# Observation of topological surface state quantum Hall effect in an intrinsic three-dimensional topological insulator

**Authors:** Yang Xu[1,2], Ireneusz Miotkowski[1], Chang Liu[3,4], Jifa Tian[1,2], Hyoungdo Nam[5], Nasser Alidoust[3,4], Jiuning Hu[2,6], Chih-Kang Shih[5], M. Zahid Hasan[3,4], Yong P. Chen[1,2,6],*

**Affiliations:**

[1] Department of Physics and Astronomy, Purdue University, West Lafayette, IN 47907 USA.

[2] Birck Nanotechnology Center, Purdue University, West Lafayette, IN 47907 USA.

[3] Joseph Henry Laboratories, Department of Physics, Princeton University, Princeton, New Jersey 08544, USA.

[4] Princeton Institute for Science and Technology of Materials, Princeton University, Princeton, New Jersey 08544, USA.

[5] Department of Physics, University of Texas at Austin, Austin, Texas 78712, USA.

[6] School of Electrical and Computer Engineering, Purdue University, West Lafayette, IN 47907 USA.

*Correspondence to: yongchen@purdue.edu

**Abstract:**
A three-dimensional (3D) topological insulator (TI) is a quantum state of matter with a gapped insulating bulk yet a conducting surface hosting topologically-protected gapless surface states. One of the most distinct electronic transport signatures predicted for such topological surface states (TSS) is a well-defined half-integer quantum Hall effect (QHE) in a magnetic field, where the surface Hall conductivities become quantized in units of $(1/2)e^2/h$ (e being the electron charge, h the Planck constant) concomitant with vanishing resistance. Here, we observe well-developed QHE arising from TSS in an intrinsic TI of $BiSbTeSe_2$. Our samples exhibit surface dominated conduction even close to room temperature, while the bulk conduction is negligible. At low temperatures and high magnetic fields perpendicular to the top and bottom surfaces, we observe well-developed integer quantized Hall plateaus, where the two parallel surfaces each contributing a half integer $e^2/h$ quantized Hall (QH) conductance, accompanied by vanishing longitudinal resistance. When the bottom surface is gated to match the top surface in carrier density, only odd integer QH plateaus are observed, representing a half-integer QHE of two degenerate Dirac gases. This system provides an excellent platform to pursue a plethora of exotic physics and novel device applications predicted for TIs, ranging from magnetic monopoles and Majorana particles to dissipationless electronics and fault-tolerant quantum computers.

**Introduction**

The integer quantum Hall effect (QHE) in a 2D electron system (2DES) under a perpendicular magnetic field is the first example of a topological quantum state[1-5]. In a QHE, the Hall conductivity ($\sigma_{xy}$, which equals to the Hall conductance) is quantized at $ve^2/h$, where the integer $v$, representing the number of filled Landau levels (LL), is a topological invariant independent of material details[3,4]. The quantized $\sigma_{xy}$ that measures this topological invariant is accompanied by vanishing longitudinal resistance ($R_{xx}$), due to dissipationless transport carried by ballistic chiral edge states, while the bulk of the sample is insulating[2]. Such quantum Hall (QH) systems have become a major paradigm in condensed matter physics, with important applications such as resistance metrology and measurements of fundamental constants[1,5]. In the past few years, it has been realized that the QHE is just one member of a much larger family of topologically distinctive quantum states, some examples of which include the quantum spin Hall (QSH) effect (also known as the "2D topological insulator") and (3D) TIs, both preserving the time reversal (TR) symmetry and induced by strong spin-orbit coupling rather than external magnetic fields (in contrast to QHE)[6,7]. Now realized or predicted in a large number of materials, TIs with their characteristic spin-helical Dirac fermion TSS have attracted intense interests for the rich physics and applications they promise.

Theories suggest that breaking the TR symmetry in TIs, either by internal magnetic doping or external magnetic fields, could result in dramatic physical consequences, in particular a *half*-quantized surface Hall conductance ($e^2/2h$), directly reflecting the topologically nontrivial nature of the underlying quantum state[8,9]. Such a half-quantized Hall conductance is also connected to the topological magnetoelectric effect (TME, formally analogous to an "axionic electrodynamics" first proposed in a particle physics context), which could lead to other novel physics such as quantized electromagnetic and optical responses (eg. Kerr rotation)[7,9]. Recent experiments have realized the predicted[10,11] quantum anomalous Hall effect (QAHE, with $\sigma_{xy}$ quantized at $2 \times e^2/2h = e^2/h$ in zero external magnetic field, with the multiplicative factor of 2 arising from two surfaces) in magnetically-doped TI films[12]. While experiments under external magnetic fields have so far reported Shubnikov-de Hass (SdH) oscillations (commonly considered a precursor for QHE[5]) from TSS in various TI materials[13-17], and even developing QHE in strained HgTe TI films[18,19], fully-developed TSS QHE characterized by well-quantized $R_{xy}$ plateaus and vanishing (dissipationless) $R_{xx}$ with truly insulating bulk has been difficult to achieve.

In this work, we report observations of well-developed TSS QHE in BiSbTeSe$_2$, an intrinsic TI material where the bulk is highly insulating with negligible conductance. The slab shaped samples with parallel top and bottom surfaces in quasi-Hall-bar configuration exhibit under high magnetic fields vanishing $R_{xx}$ with quantized $\sigma_{xy}=ve^2/h$, where $v$ is an *integer* and is equal to $(N_t+N_b+1)=(N_t+1/2)+(N_b+1/2)$, with $N_{t(b)}$ being the corresponding LL index for the top (bottom) surface (each surface filling half integer LLs). The quantized $\sigma_{xy}$ is insensitive to the sample thickness and can be understood as the sum of two half-integer quantized Hall conductances (QHC) of the two parallel conducting surfaces, even though the individual surface Hall conductance cannot be directly measured.

**Materials**

Commonly-studied "prototype" 3D TIs, nearly-stoichiometric Bi$_2$Se$_3$, Bi$_2$Te$_3$ and Sb$_2$Te$_3$, often have significant bulk conductance due to naturally-occurring defects and the resulting un-intentional bulk doping[13,14,20,21]. It has been a challenge to eliminate the bulk conduction and reveal the transport signature of the Dirac carriers in TSS[13,14,21,22]. Recently, ternary and quaternary tetradymite TI materials such as $Bi_2Te_2Se$ and $Bi_{2-x}Sb_xTe_{3-y}Se_y$ have been found to exhibit large bulk resistivity and more intrinsic TI behaviors[15,16,23-25]. We have grown (see Methods) high-quality single crystals of Bi-Sb-Te-Se solid solution with Bi:Sb:Te:Se ratio close to 1:1:1:2, referred to as $BiSbTeSe_2$ (or simply abbreviated as BSTS) in this paper. Various material characterizations have been performed on our BSTS crystals (see Supplemental Figs. S1-S3), including the confirmation of the TI nature by the angle-resolved photoemission spectroscopy (ARPES, Fig. S2) revealing the Dirac TSS band with a well-isolated Dirac point (DP) located close to the Fermi level ($E_F$), and measurement of a large bulk bandgap of ~0.3 eV (Fig. S3, by scanning tunneling spectroscopy), all in good agreement with previous measurements in this material system[23,26]. Hall coefficients measured in relatively thick crystals (Fig. S4) at low temperature indicate the bulk carrier density to be negligibly small (on the order or below the extracted effective 3D carrier density of ~$10^{15}$ cm$^{-3}$ that mostly come from the surface) and at least an order of magnitude lower than the lowest value reported previously[13,16,17,25]. Typical surface Hall mobilities range from 1000 to 3000 cm$^2$/Vs in our samples.

**Results and Analysis**

We first discuss transport measurements performed at zero magnetic field ($B$=0 T) on BSTS samples with various thicknesses ($t$). Fig. 1a shows the temperature ($T$) dependence of the sheet resistance $R_{sh}$ in 5 selected samples. The thicker samples (eg., $t$=37 μm) exhibit an insulating behavior (attributed to the bulk) at high $T$ and a saturating and weakly metallic behavior (attributed to the residual metallic surface conduction after bulk carriers freeze out) at low $T$. As the thickness (thus the bulk portion) is reduced (eg. $t$=160 and 480 nm), the bulk insulating behavior weakens and moves to higher $T$, while the $T$ range for the low-$T$ metallic conduction expands. For thinnest samples ($t$=20 nm), $R_{sh}$ is relatively $T$-insensitive and metallic over the entire $T$ range measured. We also note that while $R_{sh}$ (attributed to the bulk) at high $T$ (>~250 K) is strongly thickness-dependent (from <10 Ω to >2 kΩ), $R_{sh}$ (attributed to the surface) at low $T$ (<50 K) reaches comparable values (a few kΩ) for all the samples despite the 3 orders of magnitude variation in their thickness. This contrasting behavior is more clearly seen in Fig. 1b, where we plot both $R_{sh}$ (left axis) as well as the corresponding 3D resistivity $\rho_{3D}$ (=$R_{sh}*t$) as functions of thickness ($t$, for all measured samples) at $T$=2 K and 290 K (room temperature). At low $T$=2 K, $R_{sh}$ is insensitive to the thickness (ranging from 20 nm to 52 μm), consistent with the conduction being dominated by the 2D surface (thus independent with thickness) with little contribution from the bulk. In this case, the resistivity ($\rho_{3D}$) normally used to characterize a 3D conducting material is no longer well-defined (and becomes strongly thickness-dependent, as seen in our data, with $\rho_{3D}$ approximately proportional to $t$ as indicated by the dotted line). On the other hand, at high $T$=290 K and for samples thicker than ~100 nm, the observed behavior becomes well described by that expected of a 3D bulk conductor ($R_{sh} \propto 1/t$ and $\rho_{3D}$~constant, highlighted by the dashed line and horizontal band respectively), consistent with the conduction now being dominated by the thermally-activated carriers in the bulk. Interestingly, even at such high $T$ (room temperature), the thinner samples ($t$<100 nm) still deviate from the 3D behavior, indicating substantial conduction contribution from the surface (see also Fig. S5).

In order to more quantitatively extract the surface contribution to the total conductance ($G^{tot}$), we fit our $R_{sh}(T)$ data to a simple model used in Ref. 27, where the total (sheet) conductance $G^{tot}(T)$=1/$R_{sh}(T)$ is the parallel sum of a thermally activated bulk conductance $G_b(T)$=$t*(\rho_{b0}e^{\Delta/kT})^{-1}$,

where $k$ is the Boltzmann constant and the fitting parameters $\rho_{b0}$ is the high temperature bulk resistivity and $\Delta$ is the activation energy, and a metallic surface conductance $G^{sur}(T)=(R_{sh0}+AT)^{-1}$, where fitting parameters $R_{sh0}$ represents a low-$T$ residual resistance (due to impurity scattering) and $A$ reflects the electron-phonon scattering[27,28]. We find that this model gives reasonable good fits to most of our data, and extract the surface conductance $G^{sur}$ for samples with different thickness (see Fig. S6 for more details). Fig. 1c shows the ratio ($G^{sur}/G^{tot}$) between the fitted $G^{sur}$ and the *measured* total conductance $G^{tot}$ as a function of thickness ($t$) at 3 representative temperatures. At low $T$ (50 K or below), the surface contribution to the total conductance is nearly 100% for all the samples (with thickness from <100 nm to >10 μm). Remarkably, in thinner samples ($t<$ 100 nm) the surface conduction can exceed the bulk conduction with $G^{sur}/G^{tot}>$60% even at room temperature ($T$=290 K). For each fitting parameter, the values for all samples (with widely-varying thickness) are found to be on the similar order of magnitude (Fig. S6, given moderate density and mobility variations expected between different samples). The averaged fitting parameters (between all samples) $\rho_{b0}$=(3.5±0.8)×10$^{-3}$ Ωcm, $\Delta$=57±4 meV, $R_{sh0}$=2.0±0.3 kΩ, $A$=6±2 Ω/K can be taken as representative material parameters for our BSTS to predict $G^{sur}/G^{tot}$ for any given thickness and temperature, as shown in the 2D color plot in Fig. 2d. We see that surface-dominant transport with ($G^{sur}/G^{tot}>$80%) can be achieved in moderate low $T$ (<50 K) for samples up to macroscopic thickness of 1 mm, or in thin (tens of nm) samples up to room temperature (300 K) in this material system. The extracted bulk thermal activation energy $\Delta$ is larger than $2kT$ for $T$=300 K, which enables surface-dominant transport at room temperature. We also note $\Delta$ is much smaller than the bulk band gap of ~300 meV. This has been noted in previous measurements in similar materials[17], and may be a result of impurity states in the bulk gap in the *compensation-doped* sample (while spatial variation in the doping can further reduce the observed $\Delta$)[29].

Measurements of the quantum Hall effects (QHE) were performed in two samples ("A" and "B") with intermediate $t$ = 160 nm and 80 nm, respectively. They are thick enough such that our Si/SiO$_2$ backgate can be used to independently[22] tune the bottom surface carrier density $n_b$ without affecting the top surface density $n_t$ (see Fig. 2a inset for device schematic), yet still thin (and small) enough to ensure negligible bulk and side surface conduction and a high degree of sample uniformity. Fig. 2a shows a representative backgate modulation of $R_{xx}$ (field effect) measured at $T$=0.35 K without external magnetic field in sample A (image shown in Fig. 2b inset). The peak in $R_{xx}$ at $V_D$~-60 V is identified as the charge neutrality point (DP) of the bottom surface, whose carriers change from electrons at higher $V_{bg}$ to holes at lower $V_{bg}$. Repeated gate sweeps can have a slight hysteresis (of several volts). The ungated surfaces (both bottom surface, as seen by the field effect, and top surface, see discussions below) are generally found to be n-type (with electron carriers) in the exfoliated thin flakes used in our QHE measurements.

Fig. 2b presents the longitudinal resistance $R_{xx}$ (=$V_{23}/I_{14}$, where subscripts label voltage ($V$) or current ($I$) leads shown in the inset) and Hall resistance $R_{xy}$ (=$V_{53}/I_{14}$) as functions of the backgate voltage $V_{bg}$, measured at $T$=0.35 K in sample A and with a magnetic field $B$=31 T applied perpendicular to the top and bottom surfaces. At the electron side ($V_{bg}>$-60 V) of the bottom surface (same carrier type as the ungated top surface), we observe well-defined quantized plateaus in $R_{xy}$ with values ~$h/e^2$~25.8 kΩ and ~$h/(2e^2)$~12.9 kΩ, accompanied by *vanishing* $R_{xx}$, over broad ranges in $V_{bg}$ around -45 V and -20 V respectively. These are the hallmarks of well-developed QHE, associated with total (top+bottom surfaces) Landau filling factors $\nu$=1 and 2 respectively (see more discussions below). In addition, we also observe a developing $R_{xy}$ plateau ~$h/(3e^2)$, accompanied by a minimum in $R_{xx}$. To gain more insights on the QHE, we perform tensor inversion to extract the 2D longitudinal and Hall conductivities $\sigma_{xx}$ and $\sigma_{xy}$, shown in Fig. 2c in units of $e^2/h$. We again observe quantized (and developing) plateaus in $\sigma_{xy}$ at $\nu e^2/h$ with integer $\nu$=1, 2 (and 3), concomitant with vanishing $\sigma_{xx}$ (minimum in $\sigma_{xx}$). Such *integer* quantized Hall conductivities (QHC) measured can be understood as the sum of a half integer QHC from the top surface (fixed at $(1/2)e^2/h$ at $B$=31 T) and another half integer QHC from the bottom surface (at $\nu_b e^2/h$ with $\nu_b$=1/2, 3/2, and 5/2 etc., depending on the $n_b$ tuned by the back gate). In other words, the observed QHC $\sigma_{xy}^{total}= \nu e^2/h = \sigma_{xy}^{top} + \sigma_{xy}^{bottom} = (\nu_t + \nu_b)e^2/h = (N_t+N_b+1)e^2/h$, with top (bottom) surface QHC $\sigma_{xy}^{top(bottom)} = \nu_{t(b)} e^2/h = (N_{t(b)}+1/2)e^2/h$, where $\nu_{t(b)} = N_{t(b)}+1/2$ and $N_{t(b)}$ are the Landau filling factor and Landau level index of top (bottom) surface corresponding to the QH state (in our case $\nu_t$=1/2 and $N_t$=0, thus $\sigma_{xy}^{top} = (1/2)e^2/h$), even though the individual (half-integer) QHC of each surface cannot be directly measured in the experiment (where we always measure the *total* Hall conductance of the two parallel conducting surfaces). Such a half-integer QHE reflecting the contribution in units of *half* quantum conductance ($e^2/2h$) from each surface is the hallmark of the topological surface state QHE unique to 3D TI, and is the main focus of this paper. Also of interest is the observation when the bottom surface is gated ($V_{bg}$<-60 V) to the hole side to have the opposite carrier type with the top surface, that weakly-developed plateau-like features appear in $\sigma_{xy}$ around $0e^2/h$ and $-1e^2/h$, accompanied by inflection points in $\sigma_{xx}$ (but no obvious corresponding features in $R_{xx}$ and $R_{xy}$, Fig. 2b). It remains to be understood whether these features may relate to some underlying QH states with $(\nu_t, \nu_b)$ =(1/2, -1/2) and (1/2, -3/2), respectively. We also reproduce all the observed QH states under $B$=-31T (Fig. S7), highlighted by the $-\sigma_{xy}$ shown in Fig. 2c, exhibiting a more "smooth" transition between the "1/2-1/2" feature and "1/2+1/2" plateau (the detailed "transition" behaviors in magnetotransport near bottom surface DP ($\nu_b$ ~0) are $B$-direction dependent and not the focus of this discussion, see also Fig. S8).

We further studied $\sigma_{xx}(V_{bg})$ at a series of magnetic fields ($B$), and found quantum Hall features can be observed down to $B$=13 T (data traces of $\sigma_{xx}$ and $\sigma_{xy}$ as well as $R_{xx}$ and $R_{xy}$ are shown in Fig. S9). Fig. 2d plots $\sigma_{xx}$ (color scale) as functions of both $B$ and $V_{bg}$. Dashed lines trace the gate voltage ($V_{bg}$) positions for the observed QH minima in $\sigma_{xx}$ and are labeled by the corresponding half integer fillings ($\nu_b$) of the bottom surface LL. When extrapolated to $B$=0 T, these lines converge to ~-58 V, the bottom surface DP ($V_D$). Dotted lines trace the maxima in $\sigma_{xx}$, labeled by corresponding integer $\nu_b$, also extrapolate and converge to the same DP. The odd integer ratio (-3:-1:1:3:5) in the $V_{bg}$ positions for the QH minima in $\sigma_{xx}$ as measured away from DP ($V_{bg}-V_D$, proportional to $n_b$), exemplified in the inset for the 5 QH minima in $\sigma_{xx}$ at $B$=31 T (corresponding to the intercepts of dashed lines on the top axis in Fig. 2d), is in contrast to the QHE for conventional 2DES, where the density (filling factor) ratio for successive QHE states ($\sigma_{xx}$ minima) would be consecutive integers. This half-integer shift in our data (integer/half-integer fillings for maxima/minima) is a consequence of the unique $0^{th}$ LL shared half-half between Dirac electrons and holes, and a manifestation of the Berry's phase $\pi$ associated with spin-helical carriers of TSS. It is known that $\sigma_{xx}$ peaks when the gate-tunable chemical potential ($\mu_b$) passes through the center of last-filled LLs where the density of state (DOS) peaks and underlying state may become delocalized[2,5]. The prominent central peak (the central line with $\nu_b$ =0, non-dispersive with $B$) at DP in our $\sigma_{xx}$ data represents the $0^{th}$ LL, that is fixed at zero energy and non-dispersive with $B$, and a hallmark of Dirac fermions.

We also measure QHE by sweeping the magnetic field ($B$) at fixed carrier densities ($V_{bg}$). Fig. 3a upper (lower) shows $R_{xy}$ ($R_{xx}$) measured *vs.* $B$ at $T$=0.35 K in sample A for various $V_{bg}$'s. The $R_{xy}(B)$ trace for $V_{bg}$=-52 V (close to bottom surface DP) exhibits a wide quantized plateau ~ $h/e^2$ for $B$>~13 T, indicating both surfaces are in the $1/2(e^2/h)$ QH state ($0^{th}$ LL, $N_t=N_b=0$, $v$=1/2+1/2=1). When $V_{bg}$ is increased to -41 V, the higher $n_b$ shifts the $v_b$=1/2 ($N_b$=0) QH state of bottom surface, thus the observed $h/e^2$ ($v$=1) plateau to higher $B$. Meanwhile, another QH plateau develops at $h/2e^2$ at lower $B$~15 T (but still >~13 T with $N_t$=0), corresponding to $N_b$=1, $v= v_t+ v_b$ =1/2+3/2=2. This QH state also shifts to higher $B$ as $V_{bg}$ increases (-35 V, -30 V and -20 V). Further increasing $V_{bg}$ to -14 V, -5 V and 2 V, yet another QH plateau develops at $h/3e^2$, now corresponding to $N_t$=0, $N_b$=2, $v= v_t+ v_b$ =1/2+5/2=3. All these QH plateaus are accompanied by vanishing $R_{xx}$ or dips in $R_{xx}$, with each QH state generally better developed at higher $B$ (reflecting the larger LL energy gap). From the evolution of the QH states seen in Fig. 3a for $B$>~13 T, we can extract the bottom surface LL index $N_b$ (note $N_t$ is fixed at 0) and perform the standard fan diagram analysis (Fig. 3b, showing $N_b$ vs $1/B$), where integer/half-integer $N_b$ is assigned the minima/maxima in $R_{xx}$ (or alternatively in $dR_{xy}/dB$, with an example shown in Fig. 3b inset). We find $N_b$ vs $1/B$ (exhibiting good linear fit, see more analysis in Fig. S10) extrapolates (in the limit of $1/B \to 0$) to an $N$-axis intercept of ~-1/2 for all $V_{bg}$'s studied. Such a 1/2-intercept in LL index is again a hallmark of Dirac fermions and underlies the half-integer QHE.

Also of interest in Fig. 3a is the trace measured with $V_{bg}$=-45 V, which is near the center of the $v$=1 QH plateau in $R_{xy}$ ($V_{bg}$) in Fig. 2b, with estimated $n_b \sim n_t$. With the two surface densities nearly matched, $R_{xy}$ exhibits a developing QH state ~$h/3e^2$ (corresponding to $N_t=N_b=1$, $v$=3/2+3/2=3) at lower $B$~10 T, in addition to the $h/e^2$ QH plateau ($N_t=N_b=0$, $v$=1/2+1/2=1) at high B. These two QH states can be considered as belonging to a half-integer QHE series of $v=g(N+1/2)$, where $N$ is the common LL index (=0 and 1 in our case) and $g$=2 represents the 2 (nearly degenerate) surfaces. Another example of such QHE with matched surface densities is shown in Fig. 2d, measured with $V_{bg}$=-28 V in a different cool-down of Sample A that resulted in higher as-cooled surface carrier densities. The QH plateau near $h/(3e^2)$ appears at higher $B$~15 T and is better developed, accompanied by a clear minimum in $R_{xx}$. The conspicuous missing of the $v$=2 QH state is again consistent with the "half-integer" QHE expected for two-flavor (representing 2 surfaces) Dirac fermions at odd-integer total filling $v=2(N+1/2)=2N+1$, with LL index $N=N_t=N_b$. The inset of Fig. 3c shows the corresponding Landau fan diagram for the common LL index $N=N_{t,b}$, with the linear fit again yielding the expected intercept of $N$~-0.5 for the Dirac fermions in high-$B$ limit (thus $v=N+1/2 \to 0$). The gate-tuned QHE from this cool down can be found in Fig. S11.

Fig. 4a shows backgate-tuned QHE ($R_{xx}$ and $R_{xy}$ vs $V_{bg}$) measured a fixed $B$=31 T in Sample B (thickness $t$=80 nm) at different temperatures ($T$). At low $T$ (4.5 K), we observe QH states corresponding to $v$=1 (well-developed) and $v$=2 (developing), interpreted (similarly to Sample A, Fig. 2) to $v_t+ v_b$ =1/2 +1/2 and 1/2 +3/2 respectively. The insensitivity of the quantization values to the sample thickness (changing by factor of 2 between samples A and B) confirm that the observed QHE arises from the surface of the intrinsic TI, and is not related to the "bulk" QHE observed previously (in highly-bulk-doped $Bi_2Se_3$, attributed to electronic decoupling between bulk quintuple layers with high Se vacancies)[30,31]. As seen in Fig. 4a, increasing $T$ does not weaken substantially the QH states till at least ~10 K, and QH states are still observable up to 35 K, before disappearing above ~50 K. Fig.4b, c show more quantitative analysis of the $T$-dependence of the QHE at $v$=1 (both surface filling $0^{th}$ LL). The $\sigma_{xx}$ (QH minimum) exhibits an thermally activated behavior ($\propto \exp(-E_0/kT)$) at higher $T$ (Fig. 4b) with an extracted activation gap $E_0$~5.6±0.3 meV, but drops more slowly with decreasing $T$ at lower $T$. The $E_0$ value is notably smaller than the theoretical energy separation $v_F(2eB\hbar)^{1/2}$~60 meV between the 0th and 1st LLs of ideal 2D Dirac fermions, for a typical Fermi velocity $v_F$ ~3×10$^5$ m/s estimated from ARPES measurements (Refs.23, 26 and Fig. S2). This could stem from both disorder-induced Landau level broadening (such an effect was also seen in graphene with similar mobility[32]) and a finite thickness (side surface) effect[33]. At even higher temperature (>50 K), the bulk conduction starts to become significant and possibly suppress the QH states. The thermally induced change $\Delta R_{xy}$ (=$R_{xy}(T)$- $R_{xy}(4.5K)$) vs $R_{xx}$ at $v$=1 follows an approximately linear relationship (with a slope of -0.18, Fig. 4c). The *T*-dependence behaviors observed in Fig. 4bc are quantitatively similar to those of the previously studied integer QHE in 2DES[5].

**More Discussions and Conclusion**

Well-developed "half-integer" QHE was previously observed for Dirac fermions (4-fold degenerate with 2 spins and 2 valleys) in graphene[34,35], giving QHC $\sigma_{xy}$= 4($N$+1/2)$e^2/h$. A zero-gap HgTe quantum well was found to give a 2DES with single-valley Dirac fermions (thus considered "1/2-graphene", with 2 spins), and exhibit well-defined QHE[36]. However, the observed QHC (in unit of $e^2/h$) takes consecutive integer values rather than odd integers only (as would be expected for 2-fold degenerate Dirac fermions), because the external magnetic field gives a substantial Zeeman splitting between the 2 spins (due to the large g-factor in HgTe) in addition to the orbital LLs. In contrast, a TI surface contains only 1 species of Dirac fermions (a single non-degenerate Dirac cone, sometimes referred to a "1/4-graphene"), resulting in a QHC of ($N$+1/2)$e^2/h$. Two such degenerate surfaces (as in Fig. 3) contribute in parallel to give the observed $\sigma_{xy}$= 2($N$+1/2)$e^2/h$ (odd integer plateaus). The surface (top or bottom) index in our sample can be viewed as a pseudospin, which is *not* affected by external B field. One can use a gate voltage (controlling the chemical potential difference between two surfaces, thus acting as a *fictitious* Zeeman field) to make the pseudospins degenerate (giving odd integers of QHC, Fig.3) or non-degenerate (giving consecutive integers of QHC, Fig. 2), realizing a more tunable system.

Developing QH plateaus have been observed in 3D TI of strained HgTe films, however the accompanying $R_{xx}$ minima remained well above zero[18,19]. Suggested possible causes for the less-well-developed QHE with substantial dissipation (nonzero $R_{xx}$) may include the non-chiral conducting surface states on the side surfaces, bulk conducting states or miss-alignment between top and bottom surface QH states[18,19]. The well-quantized $R_{xy}$ (and $\sigma_{xy}$) accompanied by vanishing $R_{xx}$ (and $\sigma_{xx}$) observed in our samples not only reflect the highly clean and insulating bulk and good overlap between top and bottom surface QH states, but may also indicate that our side surface TSS has been gapped out to be non-conducting (noting the quantization energy gap for a ~100 nm thick side surface is ~70 K, above the temperatures where we observe QHE; this gap can be further increased by any magnetic field component perpendicular to the side surface).

It has been pointed out that QH physics in the closed 2D surface (topologically equivalent to a sphere) of a 3D TI can present subtle aspects unprecedented in previously studied (non-TI based) QH systems. For example, interesting questions have been asked about the QH chiral edge current, believed to become chiral side surface current, which is spin-helical and shared between the two "half" QH state in top and bottom surfaces in a TI slab like our sample. Theories even differed on the observabilities of well-developed QHE in standard transport measurements[33,37-39]. Our experimental data confirm that well-defined QHE can be observed in multi-terminal TI hall bars, and will provide important insights to understand the TSS QHE.

In contrast to the well-developed TSS QHE we observed when the top and bottom surfaces have the same carrier type (electrons), the QHE features appear very different and conspicuously *not* well-developed when the two surfaces have opposite carriers (Fig. 2, see also Figs. S7 and S11) --- with only weakly developing plateau-like features in $\sigma_{xy}$ and substantial $\sigma_{xx}$ (but no obvious features in $R_{xy}$ and $R_{xx}$). This might indicate dissipation arising from two surfaces contributing opposite QH edge chiralities. Such an interesting situation has not been well addressed in previous experiments or theories, and warrants further studies to understand the fate of edge states and QH transport in this case.

The intrinsic TI studied in this work, exhibiting surface-dominated conduction at temperature as high as room temperature and surface Dirac fermion QHE at temperature as high as ~35 K, demonstrates some of the most salient hallmarks of topological surface state transport free from contamination or complications of bulk conduction. This system could provide an excellent "clean" experimental platform to pursue the plethora of exciting physics and applications proposed for ideal TIs, such as Majorana fermions and topological magnetoelectrics when interfaced with superconducting or magnetic materials respectively.

**Methods**

Bulk single crystals of nominal composition $BiSbTeSe_2$ were grown by the vertical Bridgman technique from appropriately weighted starting materials in a two-zone high-temperature furnace. The value of temperature profile in the growth zone was regulated by changing the distance between heating zones. All the syntheses were performed in closed graphitized fused quartz ampoules evacuated for several hours, and sealed off under a dynamic vacuum of ~$10^{-8}$ torr. After a preliminary reaction at 400 $^\circ$C, followed by a slow heating up to 770 $^\circ$C for approximately 6 hours, the melt was allowed to homogenize for 24 hours under high linear temperature gradient to promote mixing. The melt was solidified with a rate of about 1 mm/hour at the linear gradients of 20-50 $^\circ$C/cm.

The crystal can be easily cleaved along the plane perpendicular to c-axis, and either fabricated into bulk devices (thickness >1 μm) with multiple indium contacts, or exfoliated (using scotch tapes) into thinner flakes (below 500 nm, with thickness measured by atomic force microscopy) placed on doped Si substrate coated with 300-nm-thick $SiO_2$. The Cr/Au electrodes for the devices on exfoliated flakes are defined by standard electron-beam lithography and metal evaporation.

Transport measurements were performed either in a variable temperature insert (VTI) system, with a temperature variable between 1.6 K and 300 K in a magnetic field up to 6 T, or in a helium-3 system with base temperature down to 350 mK and magnetic field up to 31 T (in National High Magnetic Field Laboratory). Four-terminal longitudinal ($R_{xx}$) and Hall ($R_{xy}$) resistances were measured with the standard lock-in technique and low-frequency (<20 Hz) excitation current of 100 nA. The corresponding 2D resistivities $\rho_{xx}= R_{xx} (W/L)=R_{sh}$ (sheet resistance, where $W$ and $L$ are the channel width and length between the $R_{xx}$ voltage probes of our quasi-Hall-bar devices), $\rho_{xy}= R_{xy}$, and conductivities $\sigma_{xx}=\rho_{xx}/(\rho_{xx}^2+\rho_{xy}^2)$, $\sigma_{xy}=\rho_{xy}/(\rho_{xx}^2+\rho_{xy}^2)$.

**Acknowledgments:** We acknowledge support from DARPA MESO program (Grant N66001-11-1-4107). H.N. and C-K. S. acknowledges support from Welch Foundation (Grant F-1672) and ARO (Grants W911NF-09-1-0527 and W911NF-12-1-0308). High magnetic field transport measurements were performed at the National High Magnetic Field Laboratory (NHMFL), which is jointly supported by the National Science Foundation (DMR0654118) and the State of Florida. We thank E. Palm, T. Murphy, J. Jaroszynski, E. Sang, H. Cao, J. Coy and T. Wu for experimental assistance.

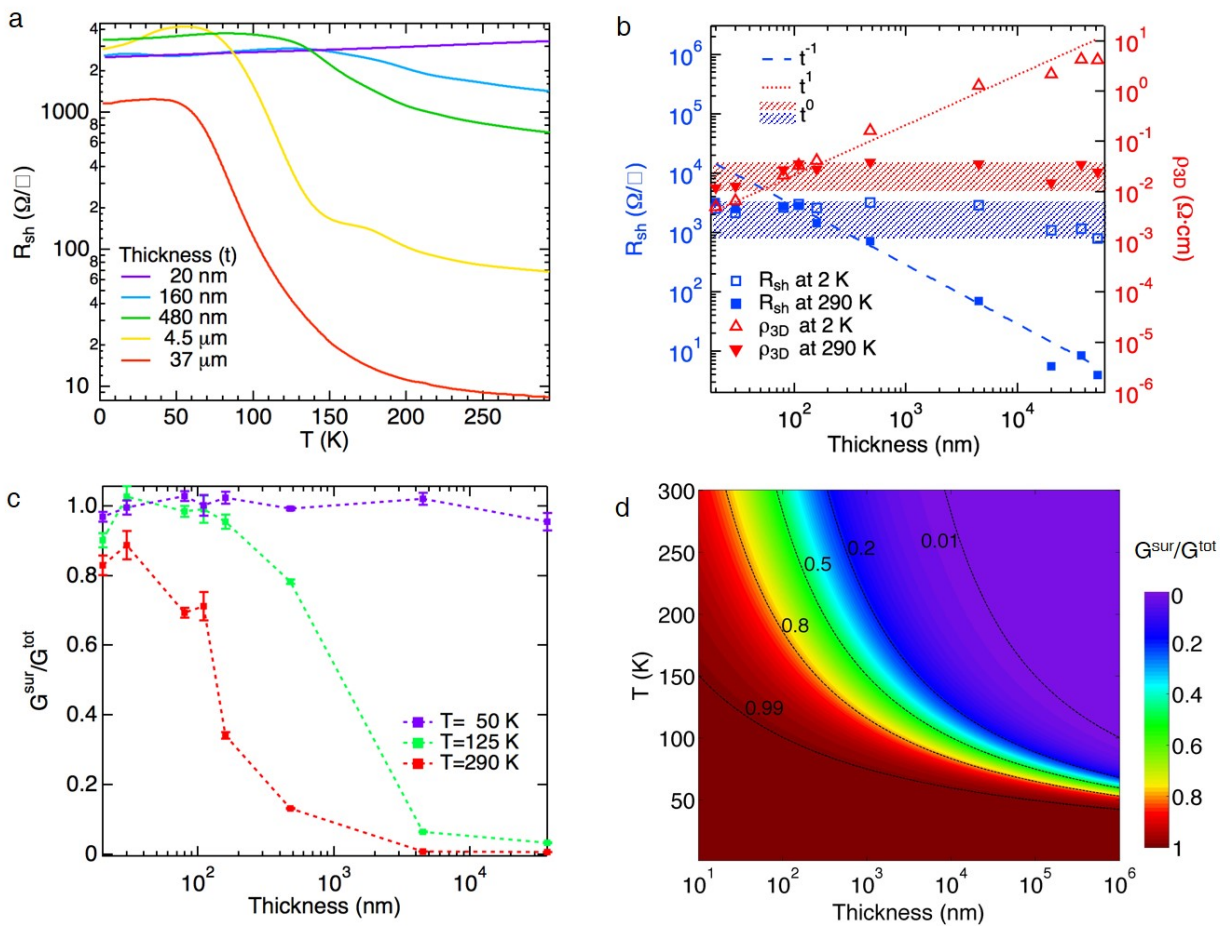

**Figure 1. Surface versus bulk conduction in BSTS.** (**a**) Sheet resistance ($R_{sh}$) measured at zero magnetic field *vs* temperature ($T$) in 5 devices with different thicknesses. (**b**) $R_{sh}$ and 3D resistivity $\rho_{3D}$ as functions of sample thickness at two typical temperatures. At $T$=2 K, the sheet resistance exhibits a 2D behavior (relatively insensitive to the thickness, highlighted by the blue horizontal band). The dashed line and red horizontal band represent expected behaviors for $R_{sh}$ and $\rho_{3D}$ of a 3D bulk conductor (our data at 290 K exhibited such 3D behavior for samples thicker than >~100 nm). (**c**) Surface to total conductance ratio ($G^{sur}/G^{tot}$) as a function of the sample thickness at three representative temperatures. Here $G^{tot}$ is the measured conductance and $G^{sur}$ is extracted from fitting the measured $R_{sh}$ vs $T$ (see text and Fig. S6 for details, the few $G^{sur}/G^{tot}$ exceeding unity by a few percent is a result of the fit slightly overestimating the measured data). (**d**) Predicted $G^{sur}/G^{tot}$ (shown in color scale) for our BSTS as a function of temperature and sample thickness. Here $G^{sur}$ and $G^{tot}$ are calculated based on the bulk and surface conductivities averaged from multiple measured devices (see text and Fig. S6 for details). Black dashed curves are contours for a few $G^{sur}/G^{tot}$ values.

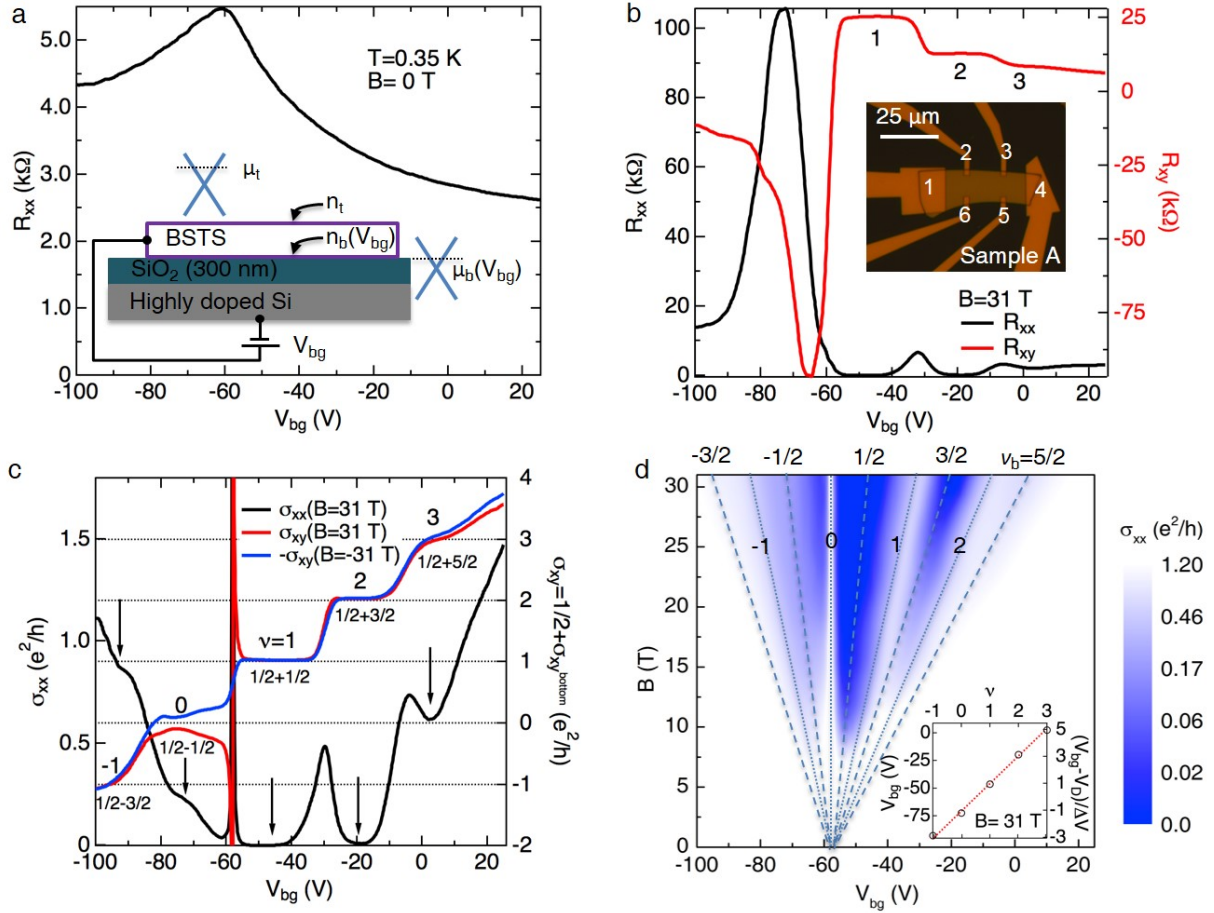

**Figure 2. Electric field effect and gate-tuned quantum Hall effect (QHE) in BSTS.** (**a**) Representative electric field effect curve, showing resistance $R_{xx}$ measured as a function of $V_{bg}$ in Sample A (a 160-nm-thick exfoliated BSTS flake). Inset is the device schematic, where the doped Si substrate (with 300 nm $SiO_2$ overlayer) is used as a backgate ($V_{bg}$ denotes the gate voltage applied) to tune the carrier density $n_b$ (thus the chemical potential $\mu_b$) of the bottom surface of the BSTS. Also depicted are the Dirac surface state bands for top and bottom surfaces, respectively. (**b**) Longitudinal resistance ($R_{xx}$) and Hall resistance ($R_{xy}$) vs backgate voltage ($V_{bg}$) measured at $B$=31T in Sample A (inset shows optical microscope image of the device) at 0.35 K. (**c**) Extracted 2D longitudinal and Hall conductivities ($\sigma_{xx}$ and $\sigma_{xy}$ at 31 T along with -$\sigma_{xy}$ at -31T), in units of $e^2/h$. Plateaus observed in $R_{xy}$ (at values of $h/ve^2$) and $\sigma_{xy}$ (at $ve^2/h$) are labeled by the corresponding *total* Landau filling factors (*v*, which are integers, with corresponding sum of top and bottom surface half integer fillings also labeled in Fig. 2c) of the quantum Hall states. Meanwhile, arrows marked the minima or inflection in $\sigma_{xx}$. (**d**) A 2D color plot showing $\sigma_{xx}$ (color scale) as a function of magnetic field (*B*) and back gate voltage ($V_{bg}$) measured at 0.35 K in sample A. A series of $\sigma_{xx}$ vs $V_{bg}$ curves (measured in steps of 2 T) are interpolated to generate this 2D plot. Dashed and dotted lines correspond to half-integer and integer Landau filling

factors (marked) for the bottom surface, respectively. Inset: the $V_{bg}$ positions for minima in $\sigma_{xx}$ (left axis) and $(V_{bg}-V_D)/\Delta V$ (right axis) as a function of ν at 31 T, where $\Delta V$ is half the averaged $V_{bg}$ separation between two consecutive Landau levels.

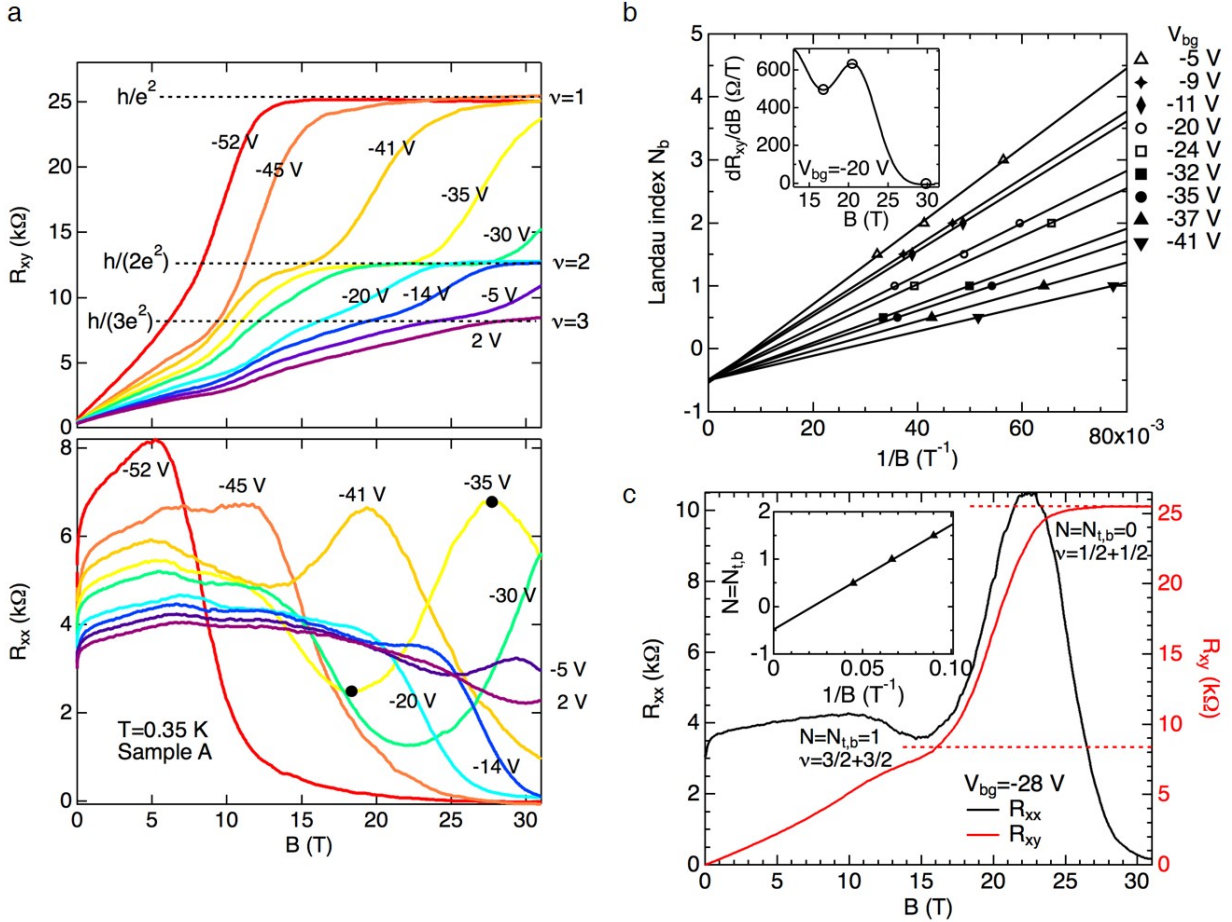

**Figure 3. Magnetic field tuned QHE. (a)** $R_{xx}$ and $R_{xy}$ as functions of B at various $V_{bg}$'s. Several observed quantum Hall plateaus are marked with corresponding total Landau filling factors ν. **(b)** Landau fan diagram for the bottom surface for various different $V_{bg}$'s. Data in solid markers are extracted from $R_{xx}(B)$ while those in empty markers from the derivative of $R_{xy}(B)$ (an example at $V_{bg}$=-20 V is shown in the inset). Examples of extracted data points are marked in the inset and $R_{xx}$ trace for $V_{bg}$=-35 V. Only data with $B$>~13 T (where the top surface is at its $0^{th}$ LL) are used. Lines denote linear fits with $N_b$-axis intercept ~ -1/2 for all datasets. **(c)** Measurements were performed in a different cool-down with different carrier densities. Magnetic field (B) dependence of $R_{xx}$ and $R_{xy}$ measured at $V_{bg}$=-28 V, where $n_b$ is tuned to match $n_t$ (top surface carrier density) such that LL index $N=N_t=N_b$. Inset shows a Landau level (LL) fan diagram, where integers N (half integers N+1/2) are assigned to index minima (maxima) in $R_{xx}$ oscillations. A linear fit gives intercept -0.5 on the N-axis. All data in **(a-c)** measured in sample A at 0.35 K.

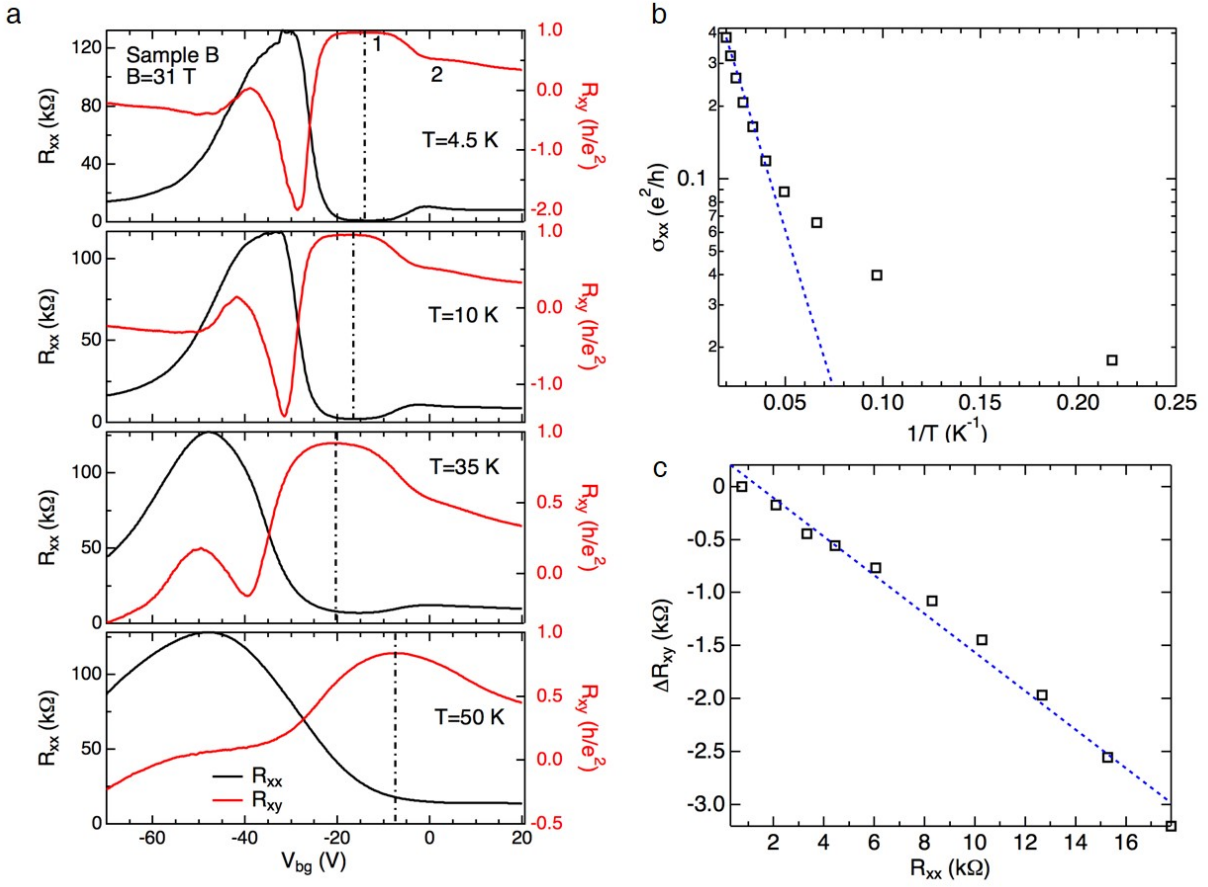

**Figure 4. QHE measured in a different sample and temperature dependence.** Measurements were performed in Sample B, an 80-nm-thick exfoliated BSTS flake. (**a**) $V_{bg}$ dependent $R_{xx}$ and $R_{xy}$ measured at $B=31$ T and 4 different temperatures (4.5 K, 10 K, 35 K, and 50 K). Vertical dot-dashed lines indicate the position of QH state at total filling factor $v=1$. (**b**) $\sigma_{xx}$ (log scale) as function of inverse temperature ($1/T$) at $v=1$. Blue dashed line indicates a fit in higher temperature with $\sigma_{xx} \propto \exp(-E_0/kT)$. (**c**) $\Delta R_{xy}$ vs $R_{xx}$ at $v=1$. Blue dashed line indicates a linear fit.

**Supplemental Materials**

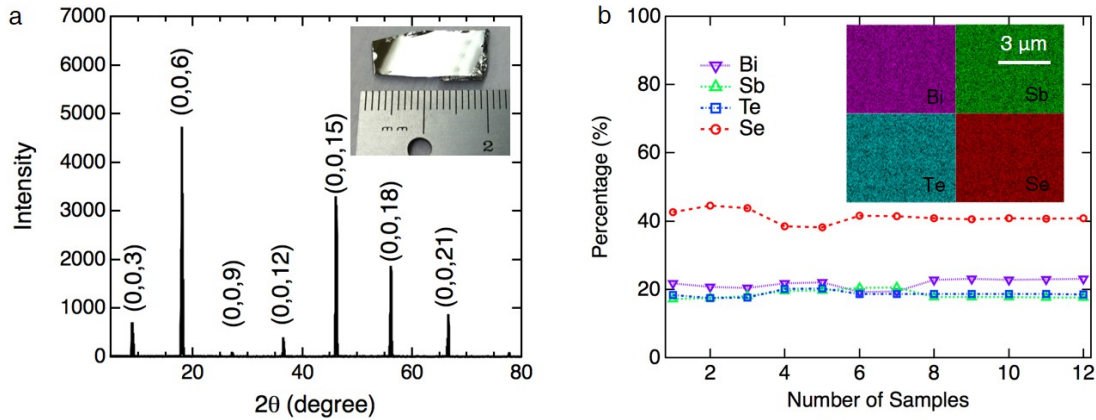

**Figure S1.** Structural characterization of BiSbTeSe$_2$ (BSTS). **a**, A X-ray diffraction (XRD) spectrum measured from a bulk crystal, in good agreement with previous XRD measurements (Ref. 24 of main text) in this material system and indicating high quality of the single crystal. The crystal was oriented with the scattering vector perpendicular to the (100) family of planes. Inset shows photo of a BSTS crystal. **b**, Molar percentage of Bi, Sb, Te, Se at 12 random positions from several pieces of BSTS crystals measured by elemental energy-dispersive X-ray spectroscopy (EDS). The EDS microanalysis was performed in an environmental scanning electron microscope (FEI Quanta 3D FEG Dual-beam SEM), operating at 10 to 15 kV with a working distance of 10 mm. The ratio of the four elements is close to the nominal stoichiometric ratio of 1:1:1:2, and shows fairly homogeneous distribution. Inset is an example of the EDS mappings in a small area.

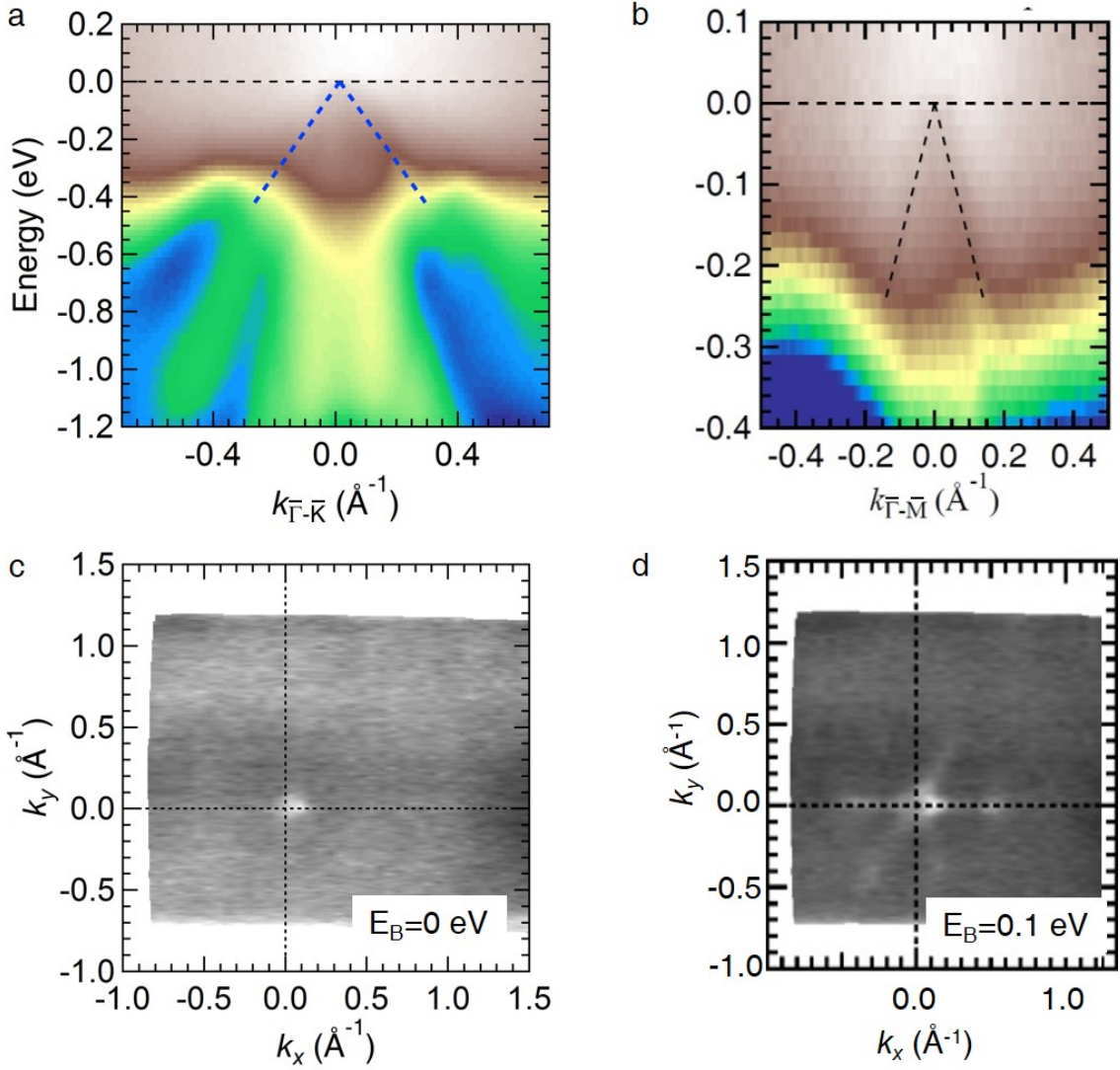

**Figure S2.** Band structure and Fermi surface of BSTS measured by angle resolved photoemission spectroscopy (ARPES). ARPES measurements were performed at Beamline 10.0.1 (HERS) of the Advanced Light Source, Berkeley, California, using a VG-Scienta R4000 electron analyzer. Energy resolution was set to ~20 meV. Samples were cleaved in situ and measured at 20 K under a vacuum condition better than $4 \times 10^{-11}$ Torr. **a**, **b**, The measured band structure (binding energy $E$ vs momentum $k$ map) of BSTS along $\bar{K}$- $\bar{\Gamma}$- $\bar{K}$ (**a**) and $\bar{M}$- $\bar{\Gamma}$- $\bar{M}$ (**b**) directions, respectively. The blue dashed lines are guides to the eye to highlight the linearly dispersive Dirac topological surface states (TSS) in the bulk band gap, with the Fermi level ($E_F$) indicated by the horizontal black dashed line. **c**, **d**, The Fermi surface map of BSTS measured at the Fermi energy ($E_F$, binding energy=0 eV, **c**) and binding energy of 0.1 eV (**d**). The point-like Fermi surface in **c** indicates that $E_F$ is located very close to the Dirac point of TSS. The star-like features in **d** are associated with bulk valence band.

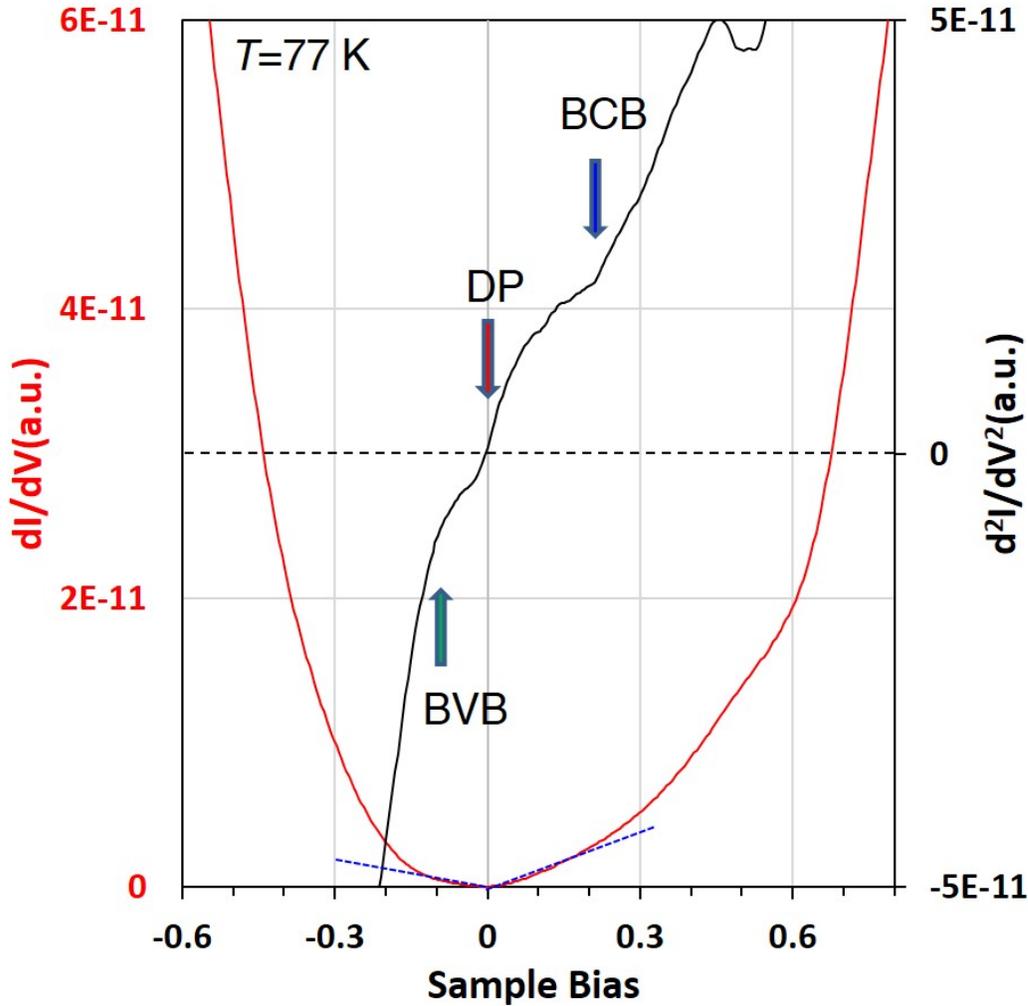

**Figure S3.** Differential conductivity d$I$/d$V$ (red) and associated d$^2I$/d$V^2$ (black) measured on our BSTS at 77 K, using STM. Both zero of d$^2I$/d$V^2$ and minimum of d$I$/d$V$ at zero bias consistently point out that the Dirac point (DP) and the Fermi level coincide, marked by a red arrow. The two dashed blue lines guide the linear dispersion of TSS, which appear as plateaus of d$^2I$/d$V^2$ curve around zero bias. The top of bulk valence band (BVB) and bottom of bulk conduction band (BCB), marked by green- and blue-arrows respectively are easier to identify from the d$^2I$/d$V^2$ spectrum. Based on the STS spectrum, we can extract a bulk band gap of ~0.3 eV and a DP-BVB separation ~0.1 eV, consistent with previous measurements by ARPES (Ref. 23 in main text).

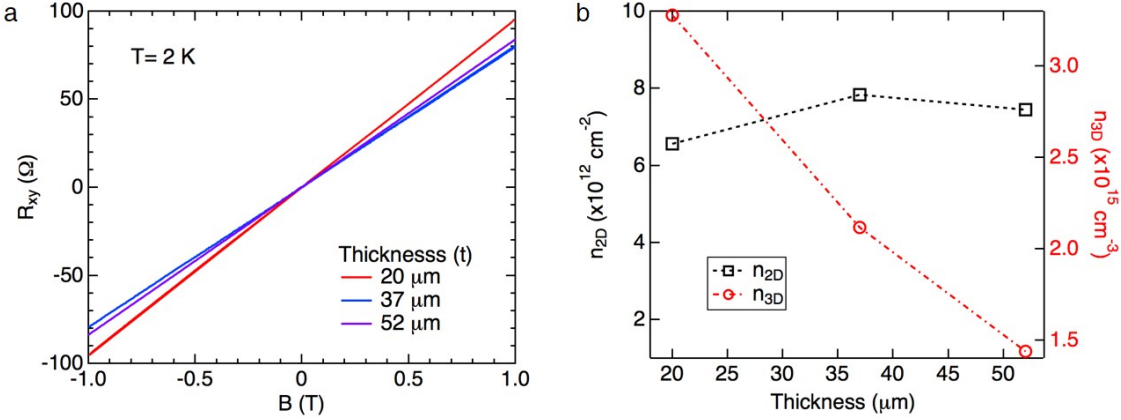

**Figure S4. a**, The Hall resistance ($R_{xy}$) at low magnetic fields in 3 different bulk samples measured at 2 K. The linear Hall slope was used to extract the 2D carrier densities to be $6.6\sim7.8\times10^{12}$ cm$^{-2}$, nearly independent of thickness (varying from 20 μm to 52 μm), indicating surface origin of the carriers (shown in **b**). In contrast, the converted 3D carrier density $n_{3D}$ (=$n_{2D}/t$) nearly scales as $t^{-1}$ and is as low as $1.4\times10^{15}$ cm$^{-3}$ for the 52-μm-thick sample. Furthermore, according to the ARPES measured band structure (ref. 23 in main text), the maximum carrier density that can be accommodated in the surface bands before occupying the bulk bands is at least $10^{13}$ cm$^{-2}$ (both surfaces combined). Therefore the measured Hall density comes mostly from the surface. This is also consistent with the surface-dominated conduction shown in Fig. 1 and the Fermi level residing inside the bulk band gap discussed above. Almost all the samples measured in our work (down to 20 nm thick) give densities on the order of $10^{12}$ cm$^{-2}$ before gating. The true bulk density should be even lower than $1.4\times10^{15}$ cm$^{-3}$, which is more than one order of magnitude lower than the lowest values from bulk-insulating 3D TIs previously reported (eg., refs. 13, 16, 17, 25 in main text).

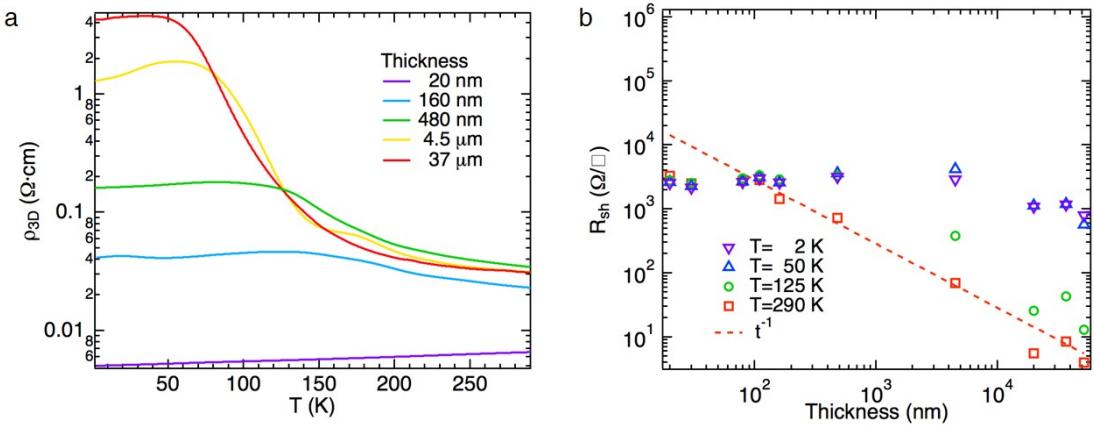

**Figure S5. a**, The 3D resistivity ($\rho_{3D}$) measured at zero magnetic field *vs* temperature (*T*) in 5 devices of different thicknesses (*t*), whose corresponding sheet resistance ($R_{sh}$) *vs T* are shown in Fig. 1a. At room temperature (290 K) $\rho_{3D}$ exhibits 3D bulk behavior (relatively independent with thickness) for most samples (except the thinnest one with *t*=20 nm, which has surface-dominant conduction even at room *T*). However, at low temperature $\rho_{3D}$ varies by three orders of magnitude (approximately proportional to *t*, as shown in Fig. 1b). **b**, $R_{sh}$ as functions of sample

thickness t at two more (intermediate) temperatures (125 K, 50 K), in addition to the data shown in Fig. 1b. $R_{sh}$ for samples with thickness below ~100 nm is relatively insensitive with temperature from 2 K to 290 K.

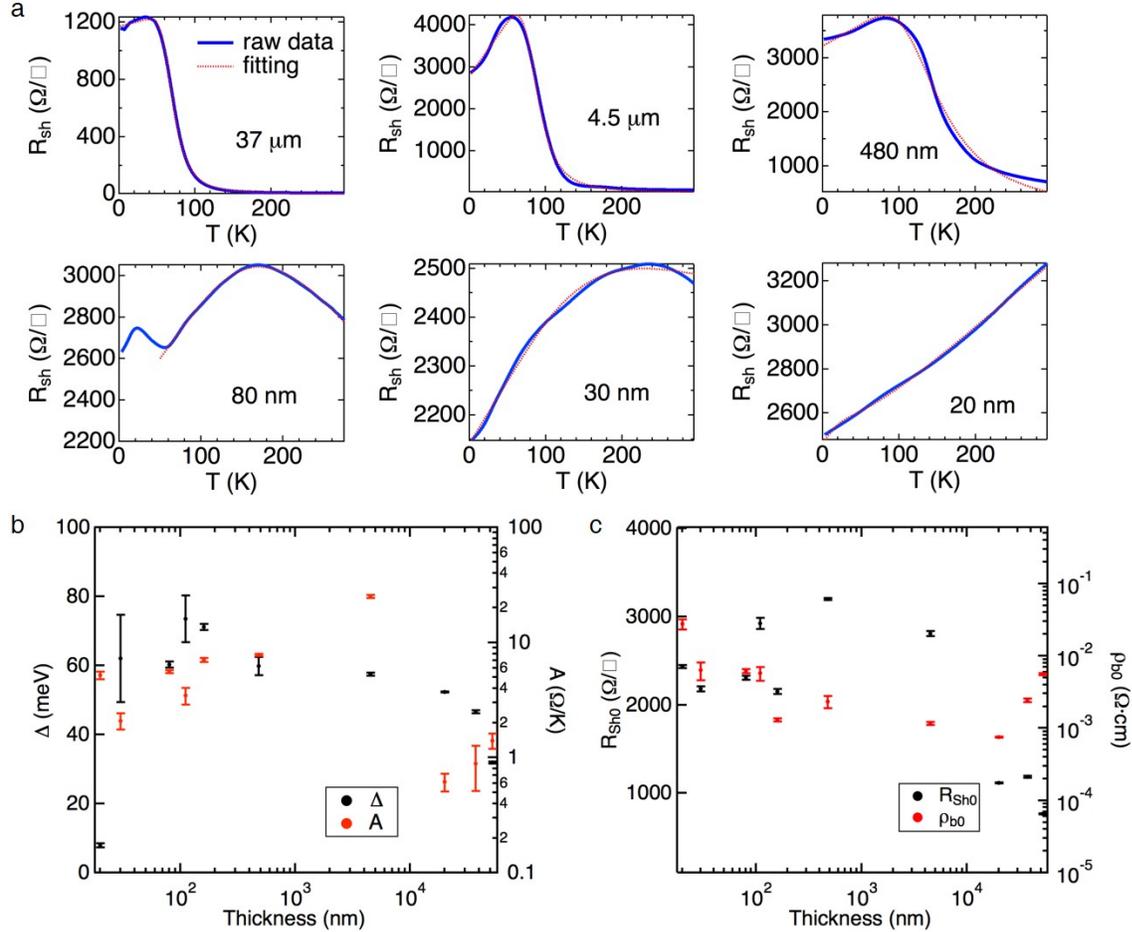

**Figure S6. a**, Fitting for $R_{sh}$ vs temperature (*T*) in 6 selected samples, using the 2 channel (metallic surface+ activated bulk) model described in the main text (following Ref. 27). This simple model fits our data remarkably well for most samples over the full temperature range. In few samples (eg. *t*=80 nm), the fitting is excellent from 300 K down to 50 K, but would not account for a small resistance peak at lower *T* (~30 K, where the fit underestimates the data by up to 10%). This peak might be due to a small part of the sample insulating with a thermal activation gap smaller than the main bulk activation gap. Each curve was fitted multiple times over different ranges of temperatures to calculate approximate confidence intervals and error bars. **b** and **c** show the fit parameters with corresponding error bars with 95% confidence level: bulk thermal activation energy *Δ*, surface electron-phonon coupling parameter *A*, low-*T* residual resistance $R_{sh0}$, high temperature bulk resistivity $\rho_{b0}$, as functions of thickness (ranging from 20 nm to 52 μm) for all the 10 samples studied. Note some fitting have small error bars barely distinguishable in the current y-axis scale. The bulk channel fitting parameters *Δ* and $\rho_{b0}$ from the 20-nm-thick sample deviate more from the others likely because the sample is too thin to accurately extract the bulk contribution. The averaged values of the parameters among all samples (excluding the 20-nm-thick sample) are used to predict $G^{sur}/G^{tot}$ for any given temperature and thickness shown in Fig. 1d.

In the fitting $A$ is a parameter describing the temperature coefficient of the surface state resistivity, reflecting electron-phonon scattering. We find that our $A$ ranges mostly from 1 to 8 Ω/K (average ~6 Ω/K), comparable with a previous measured ~3 Ω/K in $Bi_2Se_3$[S1].

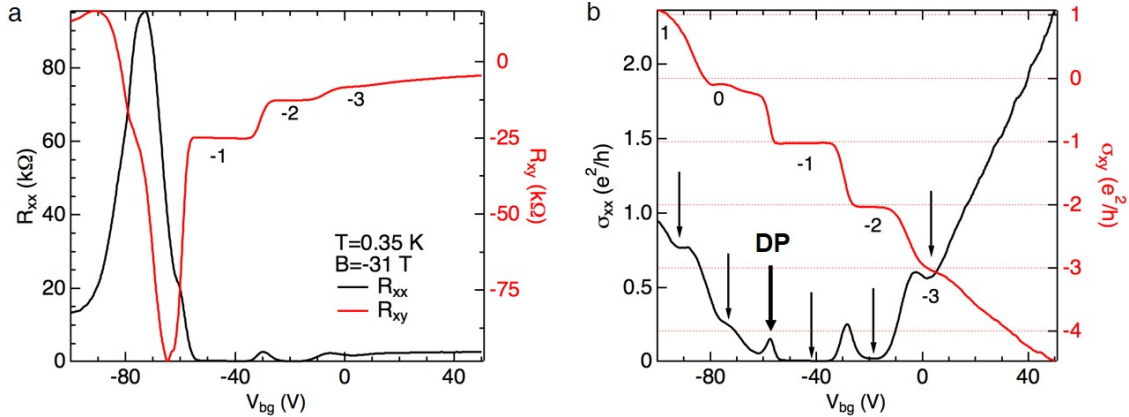

**Figure S7. a**, $R_{xy}$ and $R_{xy}$ as functions of $V_{bg}$ at $B$=-31 T for sample A, exhibiting similarly well-defined QHE as seen in Fig. 2b (for $B$=31 T). Near the bottom surface DP (away from the QH states) the $R_{xy}$ can deviate from the normal antisymmetric behavior between opposite $B$ field directions (Fig. 2b, see also Fig. S8). **b**, Corresponding $\sigma_{xy}$ and $\sigma_{xx}$ as functions of $V_{bg}$ at $B$=-31 T. Compared with the data at 31 T shown in Fig. 2c, $\sigma_{xy}$ has a smoother transition through the bottom surface DP ($V_D$ ~-60 V), and $\sigma_{xx}$ has a smaller peak (indicated by the bold arrow) at $V_D$ associated with the 0$^{th}$ LL.

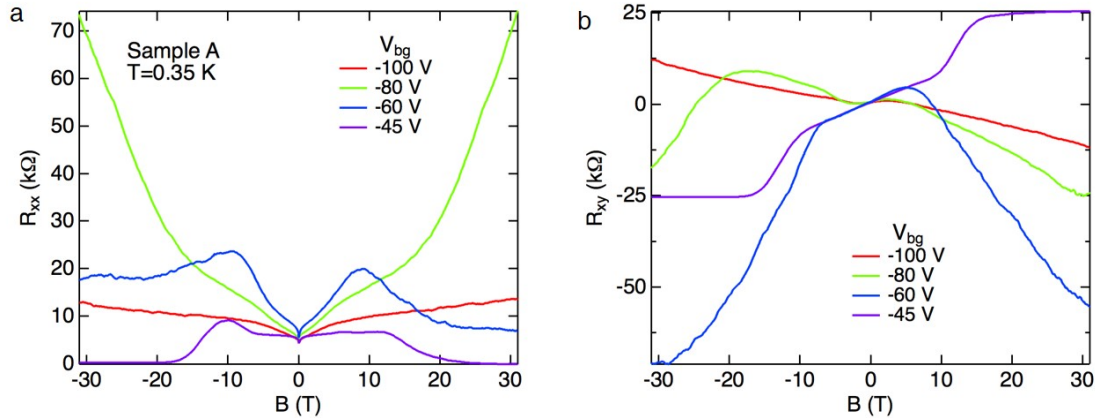

**Figure S8.** Longitudinal resistance $R_{xx}$ (**a**) and Hall resistance $R_{xy}$ (**b**) as functions of magnetic field $B$ at four representative backgate voltages in Sample A measured at 0.35 K. At $V_{bg}$ =-45 V where both the top and bottom surfaces are n-type (with electron carriers), $R_{xy}(B)$ is antisymmetric with $B$ field (as in usual Hall transport) and exhibits QHE at high $B$. When $V_{bg}$ passes the bottom surface Dirac point (~-60 V), the bottom surface starts to have opposite carriers (holes, with likely puddles of both electrons and holes near DP) from the top surface

(still electrons) and $R_{xy}(B)$ can strongly deviate from the usual antisymmetric behavior in $B$ field in this "electron-hole competing regime", while the magnetoresistance $R_{xx}$ also shows notable enhancement (a). At more negative $V_{bg}$=-100 V, $R_{xy}$ mostly recovers the antisymmetric behavior. In most of the B field range, $R_{xy}$ now takes the opposite sign from the -45 V data, indicating the transport is dominated by the bottom surface (with holes as carriers). However, at low B field, $R_{xy}$ still has the same sign as the -45 V data, reflecting the influence of the n-type top surface (see also Fig. S10a).

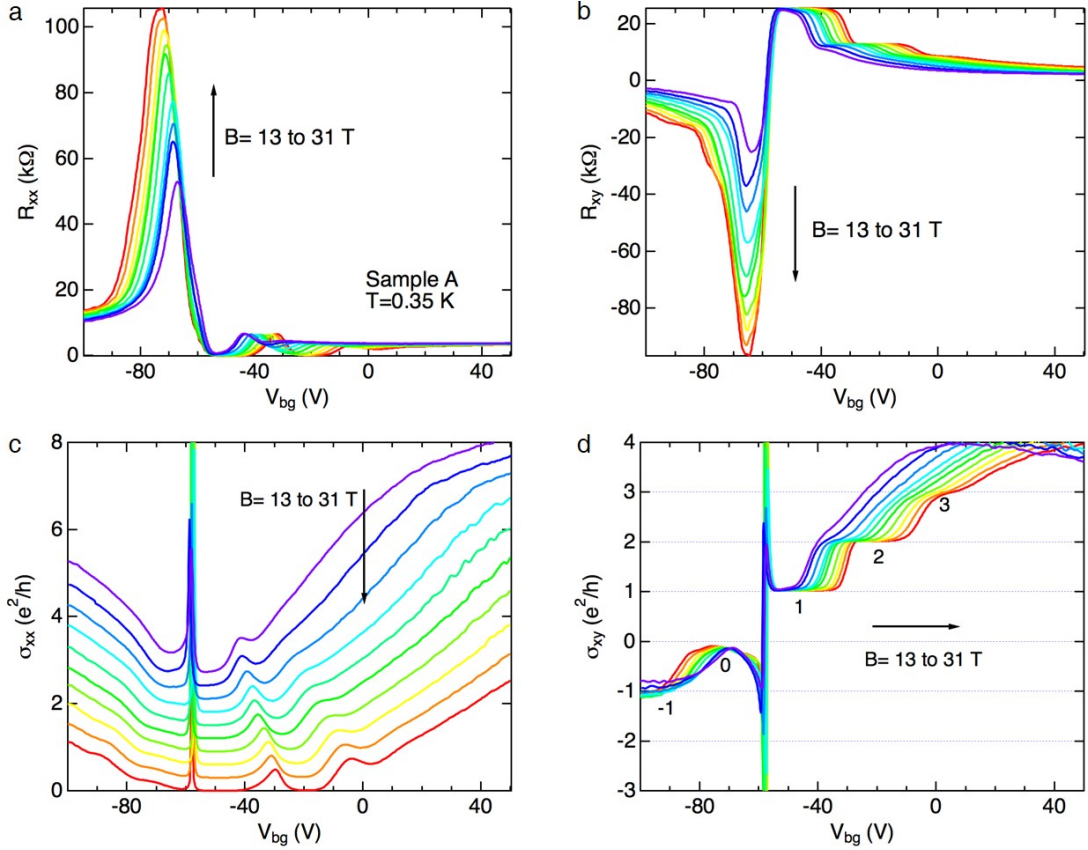

**Figure S9.** Longitudinal resistance $R_{xx}$ (**a**), Hall resistance $R_{xy}$ (**b**) and corresponding conductivities $\sigma_{xx}$ (**c**) and $\sigma_{xy}$ (**d**) as functions of $V_{bg}$ at various magnetic fields (indicated by arrows, from 13 to 31 T, in increment of 2 T) in Sample A measured at 0.35 K. In **c**, the curves are shifted vertically (in consecutive step of $0.3e^2/h$) relative to the 31 T trace for clarity. The peak in $\sigma_{xx}$ near -60 V is associated with the bottom surface Dirac point and $0^{th}$ LL (the slight fluctuation in this peak position is due to a small hysteresis in repeated gate sweeps, and has been corrected in the 2D color plot in Fig. 3d).

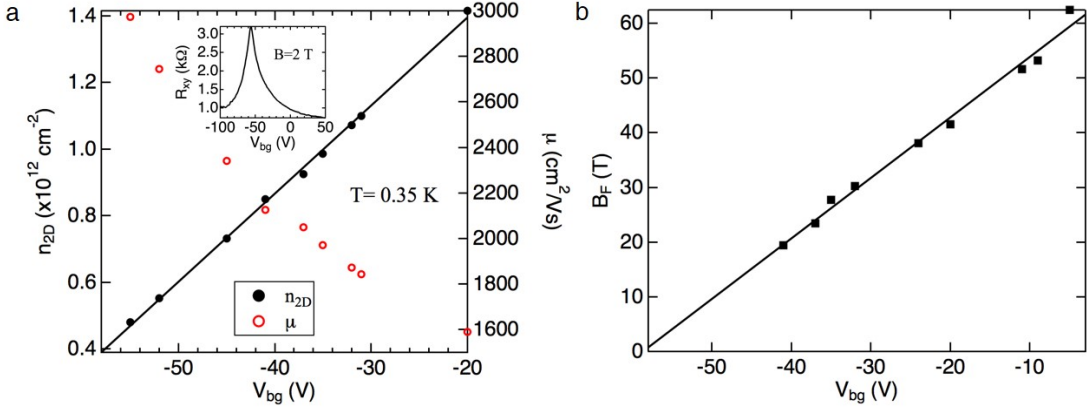

**Figure S10. a**, Total 2D carrier density $n_{2D}$ and mobility μ extracted from low-$B$ field (<~2 T) transport measurements for sample A at different $V_{bg}$'s when both surfaces have electron carriers (n-type) such that the $R_{xy}(B)$ is linear in low-$B$ regime (see Fig. 3a). The linear fit of $n_{2D}$ vs $V_{bg}$ gives a gate efficiency of $2.7 \times 10^{10}$ cm$^{-2}$/V and the extrapolation to $V_{bg}$~ -60 V (Dirac point of bottom surface) gives an approximate top surface density ~$0.4 \times 10^{12}$ cm$^{-2}$. The highest mobility ~3000 cm$^2$/Vs is extracted when $V_{bg}$ is close to bottom surface DP. Inset shows $R_{xy}$ as a function of $V_{bg}$ at $B$=2 T. As the top surface is unaffected by backgate, $R_{xy}$ remains substantially positive (corresponding to n-type carrier for the total system) at bottom surface DP. The maximum $R_{xy}$ reached close to bottom surface DP ($V_{bg}$~ -60 V) can also be used to extract the top surface density ~$0.4 \times 10^{12}$ cm$^{-2}$, consistent with the analysis above. At more negative $V_{bg}$ (<-60V), the holes from the bottom surface will compensate for the electrons from the top surface and start to *reduce $R_{xy}$*. Even at $V_{bg}$=-100 V, $R_{xy}$ remains above zero. This is consistent with the observation that the slope of low-$B$ $R_{xy}(B)$ is always positive in the measured $V_{bg}$ range (see Fig. S8b). We also note that the behaviors near bottom surface DP of $R_{xx}$ and $R_{xy}$ vs. $V_{bg}$ at low $B$ fields (Fig. 2a, Fig. S10 inset) are quite different from those at high B fields (eg., Fig. 2b, Fig. S7 and S8) where more complicate behaviors arise from the competition between electron and hole QH transport from the two surfaces. **b**, SdH oscillation frequency $B_F$ of bottom surface extracted from Fig. 3b as a function of $V_{bg}$. As the bottom surface density can be extracted as $n_b=eB_F/h$, the gate efficiency can be extracted from the linear fit of $B_F$ vs $V_{bg}$ to be $2.7 \times 10^{10}$ cm$^{-2}$/V, consistent with the result in **a**. It is notable smaller than $7.3 \times 10^{10}$ cm$^{-2}$/V given by the simple capacitance of 300-nm thick SiO$_2$, possibly due to trapped charged impurities in the oxide or other screening effects. The extrapolated linear fit to $V_{bg}$=~-60 V gives $B_F$ close to zero, consistent with expected vanishing carrier density of bottom surface near its DP. The measurements of **a** and **b** are performed at $T$=0.35 K.

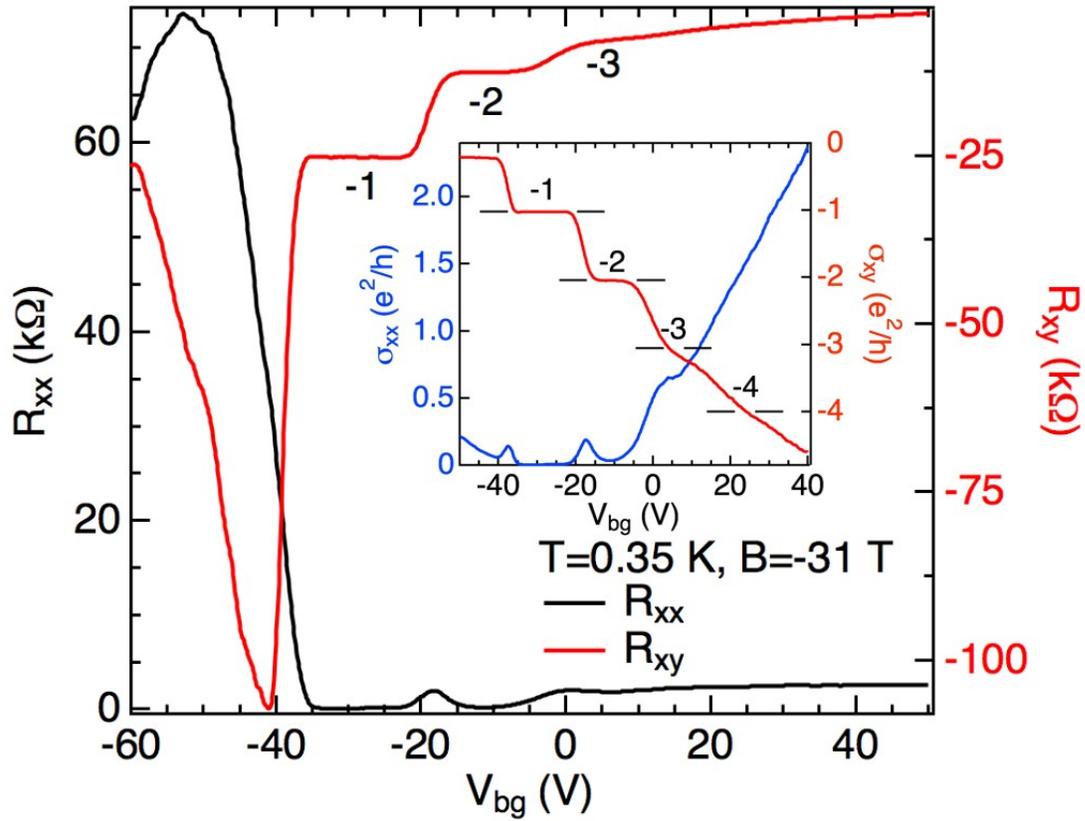

**Figure S11.** Gate tunable QHE measured in Sample A in another cool-down, with higher as-cooled densities compared to the cool down where most other data (eg. Fig. 2 and Fig. 3ab) from this sample were taken. Main panel shows $R_{xx}$ and $R_{xy}$ (and inset shows corresponding $\sigma_{xx}$ and $\sigma_{xy}$) vs $V_{bg}$ measured at $B=-31$ T and $T=0.35$ K. The top and bottom surfaces have comparable electron densities at $V_{bg}=-28$ V, where the magnetic field tuned QHE are measured and shown in Fig. 3c.